\documentclass{article}
\begin{document}



\def    \beq    {\begin{equation}} \def \eeq    {\end{equation}}
\def    \bea    {\begin{eqnarray}} \def \eea    {\end{eqnarray}}


\title{AdS$_3$ OM theory and the self-dual string or Membranes ending on the Five-brane}
\author{{\sf D. S. Berman}\footnote{berman@racah.phys.huji.ac.il} \\
{\it Racah Institute for Physics, Hebrew University,} \\
{\it Jerusalem 91904, Israel}\\[15pt]
{\sf P. Sundell}\footnote{p.sundell@phys.rug.nl}\\
{\it Institute for Theoretical Physics, University of Groningen,}
\\ {\it Nijenborgh 4, 9747 AG Groningen, The Netherlands }}

\maketitle

\begin{abstract}

We describe properties of the M-theory five-brane containing $Q$
coincident self-dual strings on its worldvolume. This is the five-brane description of Q membranes ending on the five-brane. In particular,
we consider a Maldacena-like low energy limit in the
six-dimensional worldvolume which yields a near `horizon'
description of the self-dual string using light open membranes,
i.e. OM theory, in an $AdS_3 \times S^3$ geometry. 

\end{abstract}

\section{Introduction}

The world volume description of the M-theory five-brane has been
the subject of much study, see \cite{m5,m5+,howe}. In some sense the
five-brane is the M-theory analogue of a D-brane in that open
membranes may end on five-branes \cite{Strominger:1996ac,chu}.
Just as for D-branes there exists an effective description that
describes the low energy worldvolume dynamics of the brane. For
the five-brane this has the field content of a (2,0) tensor
multiplet consisting of a self-dual two form, five scalars and
symplectic Majorana spinors. (Throughout this paper we will
restrict ourselves to a single M5 brane; the correct worldvolume
description of coincident M5 branes is not known.) The covariant
equations of motion were found using various methods in
\cite{howe}; the remarkable feature of these equations is the
non-linear self-duality constraint. This gives a non-linear,
interacting theory on the five-brane worldvolume. The
non-linearities are governed by the eleven-dimensional Planck
length, $\ell_p$, which is associated to the tension of the
M-theory branes.

There are string-like solutions to these equations \cite{sds}
that are charged with respect to the self-dual two-form. These
solutions, known as self-dual  strings, have the interpretation
of the ending of a membrane on the five-brane. As such it is the
M-theory analogue of the BIon \cite{gary}.

Recently, there have been attempts to investigate the five-brane
in a particular limit where the Plank length is taken to zero and
at the same time the background field strength on the five-brane
becomes near-critical. This leads to a new effective scale,
$\ell_6$, on the five-brane in the limit, while decoupling the
brane from the background eleven-dimensional supergravity. This
has been called open membrane (OM) theory, \cite{OM,us2,us3}. One
way of describing this limit, given in detail in \cite{us2}, is
through a conjectured effective metric on the five brane worldvolume called
the {\it{open membrane metric}}, $G^{OM}_{\mu\nu}$. This metric it
was argued, is the metric as seen by open membranes ending on the
five-brane and is the natural analogue of the open string metrics on the
D-branes. The crucial property of the OM limit is that the open
membrane metric scales in such away that the effective open
membrane tension given by $\ell_p^{-2} G^{OM}_{\mu \nu}$ is fixed. This
limit is the M-theory analog of the noncommutative open string
limit, \cite{strom}. Thus the OM theory has been conjectured to
be the UV-completion of the $(2,0)$ theory with constant background three form. As such it would
subsume all the decoupled noncommutative theories on D-branes and
NS$5$-branes \cite{OM,us2,Larsson}.

It is not certain how one should view the open memebrane metric given
that it is not derived from a fundamental theory of membranes. Being
pessimistic one might simply say that it is just a convenient way to
encode the OM limit. It seems very natural however that given there is
an open string metric such an object will lift to some sort of open
membrane metric in M-theory. (We will discuss the quantitative evidence for this metric when we introduce its precise form in the following section.)

A conceptual drawback of the above OM limit is that the
near-critical field must be switched on by hand by introducing an
an external source of delocalized (smeared) membranes, which
breaks the five-brane $ISO(5,1)$ symmetry. This raises the natural
question whether one could also create an environment with
critical field by going close to a single localized open
membrane, that is the above-mentioned self-dual string soliton.

In what follows we wish to describe the five-brane
worldvolume region near the core of the self-dual string in a low
energy, `near horizon' limit a la Maldacena \cite{malda}. In the
geometric picture where the five-brane is embedded in a flat
eleven-dimensional target space this region is a tube with the
topology ${\sf R}^2\times{\sf R}^+\times S^3$ that extends away
from the five-brane such that the radius of the transverse
three-sphere goes to zero far from the five-brane. We shall refer
to this tube region as the near horizon region. There is three-form field
strength trapped inside the tube, which turns out to be near-critical (see below (\ref{lim})). By considering the effective open membrane tension in this limit, it
turns out that the dynamics in the tube region is that of OM
theory expanded around $AdS_3\times S^3$. The germ of this idea was suggested already in \cite{us1}.

We of course expect similar results to hold for many other worldvolume
solitons, such as a $q$-brane soliton in a $p$-brane will give rise to
open $q$-brane theory \cite{OM,Larsson} in a geometry which is
conformal to $AdS_{q+1}\times S^{p-q}$ (with conformal factor playing
the role of a running brane coupling), where the anti-de Sitter space
corresponds to the directions of the worldvolume electric field
strength and the sphere to the directions of the worldvolume dual
magnetic field strength.  Related ideas concerning the fuzzy geometry
of BIons have been explored extensively in \cite{myers}. We wish to
point out that here we are moving beyond the low energy theory given
by the (2,0) tensor multiplet on the brane which becomes invalid deep in the throat as pointed out in \cite{myers} for the case of the BIon.

Of course, there is a complementary picture of this near horizon region given by the theory of Q coincident membranes with a boundary \cite{lisa}.

By examining the absorption behavior of the self-dual string we find that
the near horizon region of the self-dual string decouples from the
asymptotic region in the low energy limit.

In summary, the $AdS_3$ OM theory provides a near horizon description
of the self-dual string that decouples from the rest of the five-brane
in the prescribed limit.

The outline of the paper is as follows. In Section 2 we review
the five-brane equations of motion and their self-dual string
solution. We then describe in detail the low energy, near horizon
limit. In Section 3 we compute the self-dual string absorption
cross-section and show its vanishing in the limit. We conclude in
Section 4.

\bigskip
\section{Self-dual strings in the M5-brane and a Maldacena style limit}

The bosonic equations of motion of the five-brane in flat
eleven-dimensional spacetime are given by the scalar equation
($\mu=0,...,5; i=6,..,11$):

\beq G^{\mu\nu}\nabla^{(g)}_{\mu} \partial_{\nu} \phi^i =0
\label{se}\eeq
and the following non-linear self-duality condition:

\beq  {\sqrt{-\det g}\over 6}\epsilon_{\mu\nu\rho\sigma\lambda\tau}
{\cal H}^{\sigma\lambda\tau}={1+K\over 2}(G^{-1})_{\mu}{}^{\lambda}
{\cal H}_{\nu\rho\lambda}\ . \label{nlsd} \eeq
Here $\phi^i$ are the five transverse scalars in a static gauge,
${\cal{H}}=db$ the three-form worldvolume field strength, $g_{\mu
\nu}=\eta_{\mu\nu}+\partial_{\mu}\phi^i
\partial_{\nu}\phi^i$ the induced metric, $\nabla^{(g)}_{\mu}$ the
corresponding covariant derivative and, finally, the scalar $K$
and the tensor $G_{\mu\nu}$ are given by

\begin{eqnarray}
\label{k} K &=&\sqrt{1+{\ell_p^6 \over 24}{\cal{H}}^2}\, ,\\
&&\cr G_{\mu\nu} &=& {1+K\over 2K}\left(g_{\mu\nu}+{\ell_p^6\over
4} {\cal H}^2_{\mu\nu}\right)\ .
\end{eqnarray}
The constant $\ell_p$ is the eleven-dimensional Planck length,
which is the only parameter of the theory. Thus, the five-brane
field equations are reliable as an effective description of
M-theory dynamics only for energies smaller than $\ell_p^{-1}$.

To describe the self-dual string solution we decompose the
five-brane worldvolume coordinates $x^{\mu}$ into the $x^0, x^1$,
which are the coordinates of the string worldsheet, and the
remaining four coordinates, denoted by $y^m, m=1,...,4$, which are
coordinates of the space transverse to the string. Going to
radial coordinates we introduce $r^2= y^m y^m$. For the self-dual
string solution, the fields are functions of $r$ only.

As shown in \cite{sds} the equations of motion (\ref{se}) and
(\ref{nlsd}) can be solved by the following:

\beq \phi^6= \ell_p f \, , \quad  {\cal{H}}_{01p}= {1 \over 4}
\ell_p^{-2} \partial_p f \, \quad {\cal{H}}_{mnp} = {1 \over 4}
\ell_p^{-2} \epsilon_{mnpq}\partial_q f \ , \label{sd} \eeq
where $f$ is a harmonic function on the transverse space:

  \beq f= 1+
{{Q \ell_p^2} \over r^2}\ . \eeq
The parameter $Q$ is essentially the charge of the string as
evaluated by integrating the ${\cal H}$-flux over the $S^3$
surrounding the string. Since the field strength is self-dual one
can show that the electric and magnetic charges are equal. The
fermions and remaining overall transverse scalars
$\phi^{7,...,11}$ are zero. This solution preserves $8$ of the
$16$ five-brane worldvolume supercharges \cite{sds}.

This solution has the interpretation of a semi infinite membrane
ending on the five-brane. The energy of this configuration is
therefore infinite corresponding to the fact that the tension of
the string will be given by the tension of the membrane times the
membrane extension out of the five-brane. In order for the string
to have finite tension one regulates the solution by demanding
that the membrane is not infinite but ends on another five-brane
some finite distance, $L$, away (measured in the flat target
space metric). The solution is still good for sufficiently large L
and r. The string tension becomes:

\beq T_{string} = \ell_p ^{-3} L     \label{tension} \eeq

We will now examine the near horizon of this solution using the 
conjectured {\it{open membrane}} metric.  The tensor $G_{\mu \nu}$ is
conformally equivalent to the open membrane metric, which is
given by \cite{us2,obm, jp}:

\beq  G^{OM}_{\mu \nu}= \left({1-\sqrt{1-K^{-2}} \over K^2}\right)^{1/3} \left(g_{\mu\nu}+{\ell_p^6\over 4}
{\cal H}^2_{\mu\nu}\right) \label{gom} \eeq
This was first determined in the near-critical limit in \cite{us2} by
analyzing the scaling properties of the OM limit, and also by invoking
IIB/M theory duality. Away from the critical limit, the conformal
factor can be fixed by demanding that $G^{OM}_{\mu \nu}$ evaluated on
a probe five-brane in the background of the SUGRA dual to OM theory
\cite{us3} is independent of the three form deformation of the
solution. This is discussed at length in \cite{obm} where the idea is
based on generalising the known properties of open string metrics
\cite{swedes1}. Perhaps most naturally, one may also consider its
dimensional reduction and map it to the open string metric for the D4
brane as was also done in \cite{jp,obm}. These three different
approaches are all consistent which encourages us to believe in the
form of this open membrane metric.

We note, however, that the near horizon geometry only requires the
near-critical behavior of the conformal factor; in fact the near
horizon geometry of the self-dual string was identified up to the
conformal scale already in \cite{us1}, though the precise definition
the limit below was not appreciated at the time. A conformally related
metric was introduced in \cite{gary2} that had advantages in
simplifying the equations of motion and was used in studying a
different strong coupling limit.

The scaling limit we wish to consider is a low energy limit, where we
keep fixed the energy on the self-dual string fixed as we send $\ell_p$
to zero. In two dimensions scalar fields have dimension zero so a vacuum expectation value of such a scalar,
describing the separation of a string from the stack, is $r
T_s^{1/2}$, where $T_s$ is the tension of the self dual string as
given by (\ref{tension}). Thus, as one requires ${r /
\ell_p^{3/2}}$ to be fixed (L is obviously fixed).

We wish to compare this limit to the near horizon limit for a stack of
membranes in M-theory. The fixed energy on a stack of membranes
associated with separating off a single membrane from the stack by a
distance r is given also by ${r / \ell_p^{3/2}}$. This may be computed
either by considering the Higgsing of D$2$ branes and then going to
strong coupling and invoking $SO(8)$ invariance, or by directly
Higgsing the scalar fields on the $SO(8)$ invariant
membrane\footnote{A physical (fixed) scalar in three dimensions has
scaling dimension half. Thus the vacuum expectation value of such a
scalar corresponding to the separation of the membrane is given by
$r/\ell_p^{3/2}$}.  Moreover, from (\ref{sd}) we see that this limit
implies that the tube region close to the self-dual string, which is
the region we are interested in examining, is cut out and kept fixed
in the fixed eleven-dimensional background.  This is obvious given the
interpretation of the self-dual string as arising from the membrane.
Thus we are led to the following low energy and near horizon limit:

\beq \ell_p \rightarrow 0 \, , \qquad {r^2 \over \ell_p^3} = u \, \,
{\rm{fixed}} \label{lim} \, , \qquad Q= {\rm{fixed}} \eeq
After inserting the solution, (\ref{sd}) into the open membrane metric
(\ref{gom}) and then taking the above limit we obtain:

\beq ds^2(G^{OM})= \ell_p^2 \left( Q^{-2/3} v^2 (-dt^2 +dx^2) +
Q^{2/3} v^{-2} dv^2 +Q^{2/3} d\Omega_3^2 \right) \label{OMm}\eeq
where we have performed the trivial coordinate transformation, $v=
Q^{-1/3} u$ so as to put the line element in the canonical form for
$AdS_3 \times S^3$. Here the radii of these two spaces are equal and
given by

\beq R_{AdS} = Q^{1/3} \ell_p = R_{S^3} \eeq

We wish to analyze the energy $E(r)$ of an object
located at distance r corresponding to an open membrane excitation
scaling as $1 \over \ell_p$. It is, however, the energy $\omega$ at
infinity that we wish to associate with the energy on the self-dual
string and keep fixed. Taking into account the red-shift these two energies are
related by:

\beq \omega = E(r) \sqrt{-G^{OM}_{tt}(r)} = {\ell_p^{-2} E(r) Q^{-2/3}
r^2} = {\ell_p E(r) Q^{2/3}  u} = \, {\rm{fixed}} \, .  \eeq
Thus such energies as measured at infinity are fixed as the limit is
taken even though the energies locally at r diverge.

Examining the OM metric (\ref{OMm}), naively, one would think the
whole metric is becoming of zero size since we have $\ell_p^2$
multiplying the whole expression. However, the important property of
this limit is that the open membrane theory on the 5-brane is kept
finite as can be seen by looking at the effective open membrane
tension given by \beq \ell_p^{-2} G^{OM}_{\mu \nu} = \, \rm{fixed}  \,
, \eeq which is kept fixed in the limit but is a function of $u$. This
is just as for the usual OM theory only there the background field,
and thus the OM metric, is a constant function \cite{us2}. Here we are
performing a similar limit, sending the tension to infinity but by
going to the near horizon region of the self-dual string as described,
the field strength increases appropriately so that the scale of open
membrane excitations remain finite (though a function of $u$).
Of course we only expect to be able to believe this description when
the fields are slowly varying which would indicate having a large
$R_{AdS}$ (in plank units). This implies the description is only valid when:

\beq
Q \gg 1
\eeq

After the dust has settled, we see that in the limit described above
we are left with the OM theory on $(AdS_3 \times S^3)_Q$.  The
subscript Q denotes that the radii in plank units are proportional to
$Q^{1/3}$.

\bigskip
\section{Absorption by the self-dual string}

In this section we wish to calculate the cross section for the
absorption of massless scalars by the self-dual string in the world
volume of the M-theory five-brane. We will adopt an entirely world
volume approach similar to that of \cite{callan,others1,others2}. We
begin by writing the equation satisfied by the s-wave with energy
$\omega$, $\phi(r,t)=\phi(r)e^{i\omega t}$, of the linear fluctuations
of the four overall transverse scalars about the self-dual string, (it
is known that there are problems when one considers higher angular
momentum modes \cite{others2}, one must take care with the validity of
the linearized approximation, this is discussed in \cite{myers}):

\beq \left( \rho^{-3} {d \over d \rho}\rho^3 {d  \over d\rho} + 1 +
{R^6 \omega^6 \over \rho^6} \right) \phi(\rho) =0 \ ,\label{eom} \eeq
where $\rho= r \omega$, $R=Q^{1/3} \ell_p$. Note, as pointed out by
\cite{strom} world volume solitons have a much sharper potential than
the Coulomb type potential typical of brane solutions in supergravity;
thus this scattering is different to that of the string in six
dimensional supergravity. Nevertheless, for small $\omega R$ one may
solve this problem by matching an approximate solution in the inner
region to an approximate solution in the outer region; this follows
closely the supergravity calculation \cite{kleb}.

To approximate the inner region we change variables,  $z= (R \omega)^3
\rho^{-2}$. Then (\ref{eom}) becomes:

\beq \left( 4 {d^2 \over dz^2} + {(R \omega)^3 \over z^3} +1 \right)
\phi(z)=0\ . \eeq
This equation may be solved when $z >> R\omega$. This implies $\rho <<
R \omega$.  In this region the solution becomes simply:

\beq \phi(\rho)= A {\rm{cos}}({(R \omega)^3 \over 2\rho^2} ) + B
{\rm{sin}}({(R \omega)^3 \over 2\rho^2})\ . \eeq
The fact the fluctuation equation becomes a linear wave equation with
the coordinate change given by the excited scalar of the background
solution was noted for the D-brane analogue in \cite{callan}. A and B
are undetermined constants.

Now, consider the substitution, $\phi(\rho) = \rho^{-3/4}
\psi(\rho)$. The equation (\ref{eom}) becomes:

\beq \left( {d^2 \over d\rho^2} -{3\over 4 \rho^2} +1 + {(R \omega)^6
\over \rho^6}\right) \psi =0 \ .\eeq
One can neglect the potential term when ${(R \omega)^6 \over \rho^4}
<< 1$. This implies $\rho >> (R \omega)^{3/2}$. In this region we may
solve the equation by Bessel functions, $J(\rho)$, and Neumann
functions, $N(\rho)$, as follows:

\beq \phi(\rho)= \rho^{-1} \left( A^{\prime} J_1(\rho) +B^{\prime}
N_1(\rho) \right)\ . \eeq
Here $A^{\prime}$ and $B^{\prime}$ are undetermined constants.

We wish to calculate the ratio of fluxes between the exterior and
interior and so we need to find an overlap region that allows us to
fix $A'$ and $B'$ in terms of $A$ and $B$. There is an overlap region,
$ R\omega >> \rho >> (R \omega)^{3/2}$, when $R \omega << 1$. In this
region we can match the solutions provided:

\beq A =i A^{\prime}/2 \, ,\qquad  B = {4 \over \pi} { 1 \over (R
\omega)^3} B^{\prime} \eeq
For small $R \omega$ the ratio of fluxes of interior to asymptotic
regions is (neglecting numerical constants)

\beq {\cal{P}} \sim (R \omega)^3\ .\eeq
The formula for the absorption cross section in $d$ spatial dimensions
is:

\beq \sigma= {(2\pi)^{d-1} {\cal{P}} \over \omega^{d-1} \Omega_{d-1} }
\eeq
where $\Omega_D$ is the volume of the unit $D$-sphere. For our case,
$d=4$ giving the following (up to a numerical constant):

\beq \sigma \sim R^3 \eeq

If we return to the limit described previously, where now we can
identify $R=\ell_p Q^{1/3}$ and $\omega$ is kept fixed, we see that
$\sigma$ vanishes and the asymptotic region indeed decouples.

\bigskip
\section{Discussion}

In summary, we have defined a Maldacena-like low energy limit on the
M-theory five-brane such that the near horizon region of a self-dual
string becomes near-critical and gives rise to a $AdS_3$ phase of the
OM theory. This describes the emerging stack of Q membranes from the
five brane perspective.

Now, in the above discussion we have neglected the ambient gravity. To
decouple this brane system from gravity, one takes $\ell_p$ to zero
with ${v^2 \over \ell_p^3}$ fixed where $v$ is now the eleven
dimensional distance from membranes. A strictly rigorous analysis is
not possible since there is no supergravity solution, however, for
large $Q$, one may ignore the back reaction of the single M5 brane and
we have a decoupled $AdS_4 \times S^7$ spacetime with radius of
curvature (using the naive estimate) going like $Q^{1/6}$ with the
$AdS_3 \times S^3$ is embedded inside whose radius of curvature goes
like $Q^{1/3}$. A similar set up is described in \cite{lisa}.

By compactifying the whole system, one introduces a plethora of
various decoupled open brane theories \cite{OM,us2,Larsson} with
coupling moduli. We expect all these to give rise to similar behavior
as the one found here, though presumably the conformal invariance is
broken in most of these cases. This of course requires the
identification of the appropriate open brane metrics and coupling; as done in
 \cite{obm}.

One might imagine that we are describing simply the usual stack of Q
coincident membranes in the limit. This is not the case however
because crucially the membranes have a boundary which is ofcourse the
five-brane. They must end on the five-brane in a smooth way that also
conserves flux. It is this that essentially alters the geometry of
the membranes.

Finally, we wish to entertain the following possibility.  One may
imagine that the the low energy effective action of the self-dual
string may be determined by writing down the action corresponding to
goldstone modes of the self-dual string, or using the more powerful
superembedding formalism of \cite{emb} which was applied to the
self-dual string in \cite{chu}. As such it will be an $N=(4,4)$ SCFT in
1+1 dimensions. Actually knowing what the precise SCFT is problematic
because of how one determines the $Q$ dependence. The approaches
described above actually only gives the action when $Q=1$ but allows
one to determine the supersymmetry. This is similar to the problem of
knowing what the conformal theories are for any of the M-theory
branes, at the moment the only thing known about them is the
supersymmetry and properties determined by the assumed duality with
eleven-dimensional supergravity. It would be interesting to see if the
work of \cite{gabriele} could be useful in describing this
{\it{non-abelian}} self-dual string theory. One significant objection
to this proposal is that the solution is singular and has infinite
tension unless regulated by the arbitrary cut off L as indicated
here. Nevertheless, if such a low energy effective action exists and
could be trusted; it would indicate a low energy duality
between the two descriptions of the decoupled region.

As an aside we wish to point out that
the six-dimensional five-brane theory should not be thought of as
the $Q\rightarrow \infty$ limit of the $AdS_3$ version of the OM
theory that we have found here, just as one should not think of
the ten-dimensional type IIB string theory as the large $N$ limit
of the five-dimensional anti-de Sitter string theory. The flat
limit instead consists of uplifting to the maximally symmetric
theory in the higher dimension, where the brane charge can be set
equal to zero. In fact, in analogy with the type IIB string
theory, we expect the $AdS_3$ OM theory to be given in terms of a
curvature expansion for large $R$ with an essential singularity
at $1/R=0$. In type IIB this pole origins from the singularity in
the string tension in the curvature expansion of string theory
about the ten-dimensional flat vacuum \cite{FL}. When expanded
about the five-dimensional anti-de Sitter vacuum the string
tension is traded for the cosmological constant by letting it
absorb powers of the string coupling (dilaton). As a result the
string tension can be sent to zero at fixed Planck length and
radius (which corresponds to sending the string coupling to
zero). In the resulting tensionless phase the original
singularity thus has translated into the singularity at $1/R=0$.

One also might be worried that in fact the OM theory does not really
decouple from the background due to its thermal properties; as was
suggested for NCOS theories in \cite{eliezer}.  It was not clear there
what such a thermal analysis revealed about OM theory  as such it
remains an open question. The relation if any to the usual $AdS_3
\times S^3$, CFT coresspondence, \cite{amit} also remains undetermined.

\section{Acknowledgements}

We are thankful to S.~Deger and A.~Kaya for valuable discussions. We
would like to thank O.~Aharony, M.~Cederwall, S.~Elitzur, G.~Feretti, 
A.~Giveon, P.~Howe, H.~Larsson, B.E.W.~Nilsson, E.~Rabinovici, D.~Roest and
A.~Shomer for stimulating comments. The basic ideas exploited in this
paper took shape during conversations together with E.~Bergshoeff and
J.P.~van der Schaar. The work of D.S.B. is supported by the European
RTN network HPRN-CT-2000-00122, the BSF-American-Israeli bi-national
science foundation, the center of excellence project, ISF and the
German-Israeli bi-national Science foundation. The work of P.S. is
part of the research program of Stichting FOM, Utrecht.

\end{document}